\documentclass[12pt]{article}
\usepackage{amsmath}
\usepackage{graphicx}
\usepackage{enumerate}
\usepackage{natbib}
\usepackage{url} % not crucial - just used below for the URL 

\usepackage{amsmath}
\usepackage{amsfonts, amssymb} %Either provides Real number and other font'related stuff.
\usepackage{amsthm} % Theorem environment
\usepackage{multirow}
\usepackage{graphicx}
\usepackage{color}
\usepackage{hyperref} %For adding hyperlinks in documents
\newtheorem{theorem}{Theorem}

\usepackage{setspace}% http://ctan.org/pkg/setspace
\usepackage{etoolbox}
\AtBeginEnvironment{tabular}{\singlespacing}% Single spacing in tabular environment
\usepackage{booktabs,dcolumn}
\newcolumntype{d}[1]{D{.}{.}{#1}}

\theoremstyle{definition}
\newtheorem{definition}{Definition}[section]

\newcommand{\blind}{1}

\begin{document}
	\bibliographystyle{abbrvnat}
	
	\def\spacingset#1{\renewcommand{\baselinestretch}%
		{#1}\small\normalsize} \spacingset{1}

	%%%%%%%%%%%%%%%%%%%%%%%%%%%%%%%%%%%%%%%%%%%%%%%%%%%%%%%%%%%%%%%%%%%%%%%%%%%%%%
	
	\if1\blind
	{
		\title{\bf A Robust, Differentially Private Randomized Experiment for Evaluating Online Educational Programs With Sensitive Student Data}
		\author{Manjusha Kancharla\thanks{
				The authors gratefully acknowledge Panduan An, Mike Baiocchi, Derek Bean, Nancy Brinkerhoff, Rick Chappell, Elina Choi, Alex Covington, Mohamed Elkhouly, Tedward Erker, John Gillett, Susan Glenn, Chelsey Green, Jessi Kehe, Sean Kent, Wei-Yin Loh, Linquan Ma, Chan Park, Vivak Patel, Mitchell Paukner, Sebastian Raschka, Kris Sankaran, and Bo Yang for their help in running the experiment. The research of Hyunseung Kang was supported in part by NSF Grant DMS-1811414. }\hspace{.2cm}\\
			Department of Statistics, University of Wisconsin - Madison\\
			and \\
			Hyunseung Kang \\
			Department of Statistics, University of Wisconsin - Madison}
		\maketitle
	} \fi
	
	\if0\blind
	{
		\bigskip
		\bigskip
		\bigskip
		\begin{center}
			{\LARGE\bf A Robust, Differentially Private Randomized Experiment for Program Evaluation Involving Sensitive Participant Data}
		\end{center}
		\medskip
	} \fi
	
	\bigskip
	\begin{abstract}
		Randomized control trials (RCTs) have been the gold standard to evaluate the effectiveness of a program, policy, or treatment on an outcome of interest. However, many RCTs assume that study participants are willing to share their (potentially sensitive) data, specifically their response to treatment. This assumption, while trivial at first, is becoming difficult to satisfy in the modern era, especially in online settings where there are more regulations to protect individuals' data. The paper presents a new, simple experimental design that is differentially private, one of the strongest notions of data privacy. Also, using works on noncompliance in experimental psychology, we show that our design is robust against ``adversarial'' participants who may distrust investigators with their personal data and provide contaminated responses to intentionally bias the results of the experiment. Under our new design, we propose unbiased and asymptotically Normal estimators for the average treatment effect. We also present a doubly robust, covariate-adjusted estimator that uses pre-treatment covariates (if available) to improve efficiency. We conclude by using the proposed experimental design to evaluate the effectiveness of online statistics courses at the University of Wisconsin-Madison during the Spring 2021 semester, where many classes were online due to COVID-19.
	\end{abstract}
	
	\noindent%
	\vfill
	
	\newpage
	\spacingset{1.9} % DON'T change the spacing!
	\section{Introduction}
	\label{sec:intro}
	
	\subsection{Motivation: Balancing Privacy and Scientific Inquiry in Program Evaluation Using Randomized Experiments}
	
	The gold standard to evaluate the effectiveness of a program, policy, or treatment on an outcome of interest is a randomized control trial (RCT). In its purest form, an investigator randomly assigns each individual $i$, $i=1,\ldots,n$, to treatment, denoted as $A_i = 1$, or control, denoted as $A_i = 0$, and observes his/her outcome $Y_{i}$. Because the treatment was randomized, both treated and control groups are similar in their unobserved and observed characteristics and thus, taking the difference of the average outcomes from the two groups yields an unbiased estimate of the average treatment effect (ATE). A bit more formally, a RCT satisfies strong ignorability \citep{Rosenbaum&Rubin} and the ATE can be identified from observed data; see \citet{ImbensRubin} and \citet{HernanRobins} for textbook discussions.

	In addition to strong ignorability, one subtle, yet important assumption underlying RCTs is that after randomly assigning treatment, individuals in the study share their response/outcomes  to the investigator. This assumption, almost axiomatic in a RCT,  is becoming less plausible in the modern era, especially in online settings where there are increasing regulations to protect users' privacy. For example,  \citet{voting} ran an online randomized experiment among 61 million users of Facebook and collected their voting behaviors as the primary outcome; this experiment attracted controversy due, in part, to concerns over data privacy \citep{ethics}. In RCT-based evaluations of educational programs, investigators often collect sensitive data on students' performance, say test scores, probation status, and class rank, as their primary outcomes \citep{class1, WILSON}. 
	
	Also, from a regulatory perspective, the European Union enacted the General Data Protection Regulation (GDPR), which limits the information that websites can collect from users \citep{EU}. In 2020, California started enforcing the California Consumer Privacy Act, which like GDPR, guaranteed certain privacy rights for online user data. The main theme of this work is to explore how to guarantee individual data privacy while still being able to use RCT-based evaluation of programs, especially those from online educational settings.

	\subsection{Review of Existing Methods to Data Privacy in RCTs}
	A popular approach to data privacy in clinical trials and A/B testing in online settings is to lock up the user data after completing the experiment and only report summary statistics for scientific dissemination (e.g., \citet{WILSON}).
	%\citep{Naltrexone, PTSD, NEZHAT2008181, voting,DomesticViolence}. 
	While this approach tilts more closely to guaranteeing privacy, a major downside is that independent replication of the original analysis is difficult, if not impossible. For instance, it is generally difficult for a future, external investigator to plot, diagnose, fit competing or new models based on summary statistics from the original experiment. Relatedly, it would be difficult for future investigators to merge the original experimental data with new experimental data to build more complex models, or to boost statistical power, especially for heterogeneous treatment effects. Even if these difficulties are deemed tolerable, this approach rests on the assumption that subjects are willing to give up their personal information to the investigators in the first place, and investigators will keep the data safe, in perpetuity.
	
	Another popular approach is based on data anonymization where identifiable information is removed, aggregated, or anonymized before sharing the ``de-identified'' data to the public. Some examples include removing any protected health information (PHI) to be compliant with the Health Insurance Portability and Accountability Act (HIPAA) \citep{HIPPA}, using $\kappa$-anonymity \citep{Samarati,SweeneyKAN}, $\ell$-diversity, or $t$-closeness \citep{Machanavajjhala2006, LiLiV}. While they are an improvement over the previous approach in terms of replicability and data sharing, it has been shown that many popular de-identification methods are not sufficient to guarantee privacy. For example, \citet{Sweeney} linked de-identified patient-specific health data to voter registration records using variables such as ZIP code, birth date and gender and observed that 87\% of the U.S. population can be uniquely identified. Also, \citet{Narayanan&Shmatikov} linked the Netflix Prize dataset containing anonymized movie ratings of 500,000 Netflix subscribers to the Internet Movie Database (IMDb), allowing re-identification of users on Netflix; this led to the discontinuation of the Netflix Prize in 2010 \citep{NeilHunt.2010}.
	
	The approach to data privacy we use in this work is differential privacy \citep{Evfimievski,Dwork2006, Dwork&Smith}, specifically local differential privacy; see \citet{Duchi} and references therein. Broadly speaking, differential privacy is a mathematical definition of privacy where nearly identical statistics are computed from a dataset, say the sample mean or the p-value from a hypothesis test, regardless of whether any one individual is present or absent in the dataset; see Section \ref{dpdef} for details. Differential privacy is considered to be the strongest form of data privacy in that if an adversary were to obtain differentially private data, it is, up to a privacy loss value $\epsilon$, impossible to re-identify the individual in the data. Due to these strong privacy guarantees, differential privacy is used by Google's Chrome browser \citep{Erlingsson_2014} and Apple's mobile iOS platform \citep{Apple} to protect their users' privacy while enabling the development of novel machine learning methods and statistical analysis.
	
	\subsection{Our Contributions} \label{sec:contribute}
	Our main contribution is to propose a simple, robust RCT that guarantees local differential privacy while allowing investigators to estimate treatment effects. Specifically, similar to a typical RCT, we assume that the investigator randomly assigns treatment to individual $i$ and therefore, the treatment value is known to the investigator. But, unlike a typical RCT, we use randomized response techniques originally from \citet{Warner1965} to collect differentially private outcome data from individual $i$, denoted as $\widetilde{Y}_{i}$, instead of the sensitive, ``true'' outcome, denoted as $Y_i$. 
	That is, the investigator only sees a privatized response $\widetilde{Y}_{i}$ along with the treatment assignment $A_i$ to estimate the treatment effect on the sensitive/true outcome $Y_i$; in contrast, a typical RCT allows investigators to see both the sensitive/true outcome $Y_i$ and $A_i$ to estimate the treatment effect on the true sensitive/outcome $Y_i$.
	
	A key innovation in our proposed experimental design that distinguishes itself from a straightforward application of existing techniques in differential privacy to RCTs is that we allow the privatized response $\widetilde{Y}_i$ to be  ``adversarial.'' 
	More concretely, unbeknownst to the investigator, some participants may provide  ``adversarial'' data to further mask their identity, say by providing a completely random value as $\widetilde{Y}_{i}$ that deviates from the experimental protocol. 
	Our proposed design allows responses from such participants, which we broadly call ``cheaters,'' and even if their identity is unknown to the investigator, their responses will not harm estimation and inference of treatment effects. In relation to works in differential privacy, a cheater represents, in a loose sense, an ``imperfect'' implementation of a differentially private algorithm where a database/central entity holding the private data may not faithfully execute the privacy-preserving algorithm, say the entity added the wrong noise to the private data or it forgot to add any noise to the private data. Our work here shows how to still obtain relevant statistics of interest even if the differentially private algorithm may be imperfectly implemented. We achieve this by using a simple idea based on sample splitting and noncompliance in psychometric testing \citep{Clark&Desharnais} where we apply two slightly different differentially private algorithms on two random subgroups of participants and re-weigh the outputs from the two algorithms via inverse probability weights to remove bias arising from cheaters; see Section \ref{sec:ident} for details. Also, in relation to works in psychometric testing, our work extends \citet{Clark&Desharnais} to allow for arbitrary types of cheaters and differential privacy; see Section \ref{sec:cheat_eg} for details.
	
	Once we have data from the proposed design, we propose two consistent estimators. The first estimator is essentially a difference-in-means estimator weighted by the proportion of non-cheaters and is similar, in form, to the local average treatment effect in the noncompliance literature \citep{AngristImbensRubin}. The second estimator is a doubly robust, covariate-adjusted estimator that uses pre-treatment covariates, if available, to improve efficiency. We also compare our design to a typical RCT that collects the true, sensitive/private outcome and assess the trade-off between statistical efficiency and data privacy. 
	
	Finally, the proposed experimental design is used to evaluate online statistics courses at the University of Wisconsin-Madison. Specifically, during the Spring of 2021 when most classroom instruction went online due to COVID-19, $n = 72$ students participated in an evaluation about the impact of instructors being present in online lecture videos on learning outcomes. Similar to prior works in this area \citep{Kizilcec, Pi&Hong, WILSON, WANG2020103779}, we find that instructor-present video lectures improved students' attention among non-cheaters. Critically, unlike these prior works, the sensitive learning outcomes from the students are guaranteed to be differentially private to any investigator (including those that actually conducted the evaluation). In fact, the proposed design received approval from the Education and Social/Behavioral Science Institutional Review Board (IRB) of the University of Wisconsin - Madison to release this data to the public for future replication; the protocol met the criteria for exempt human subjects in accordance with categories 1 and 3 as defined under 45 CFR 46 \citep{exempt}.

	\section{Setup, Review, and Definition of Cheaters}\label{sec:defs}
	
	\subsection{Review: Notation, Potential Outcomes, and RCTs}\label{sec:notation}
	We review the potential outcomes notation used to define treatment effects \citep{Neyman, RubinPotOC}. For each individual $i=1,\ldots,n$, let $A_i \in \{0,1\}$ denote the binary treatment assignment with $A_{i} = 1$ denoting treatment and $A_{i} =0$ denoting control. Let $Y_{i}(a) \in \mathcal{D} \subseteq \mathbb{R}$ denote the  potential outcome of individual $i$ under treatment assignment value $a \in \{0,1\}$ and $\mathcal{D}$ denote the support of the outcome. For simplicity, we consider $\mathcal{D}$ to take on binary values, $\mathcal{D} = \{0,1\}$. In Section \ref{sec:DP}, we show that as long as $\mathcal{D}$ is restricted so that differential privacy is well-defined (see Chapter 2.3 of \citet{Dwork&Roth}), our proposed experimental design remains differentially private. Finally, let $Y_i \in \mathcal{D}$ and $X_i \in \mathbb{R}^p$ be the observed outcome and $p$ pre-treatment covariates, respectively, for individual $i$.
	
	The estimand of interest is the average treatment effect (ATE) defined as $\tau = E[Y_i(1) - Y_{i}(0)]$. To identify $\tau$, the following assumptions are usually made; here, we write the assumptions under a RCT, but the interested reader is referred to  \citet{ImbensRubin} and \citet{HernanRobins} for identification strategies outside of RCTs.
	\begin{itemize}
		\item[(A1)] Stable unit treatment value assumption \citep{Rubin}: $Y_i = Y_{i}(1) A_i + Y_{i}(0) (1 - A_i)$
		\item[(A2)] Complete randomization of treatment: $A_i \perp X_i, Y_i(1), Y_{i}(0)$
		\item[(A3)] Overlap: There exists $0 < \zeta<1$, where $1- \zeta <  \delta = \Pr(A_i = 1) < \zeta$.
	\end{itemize}
	Briefly, assumption (A1) states that there are no different versions of the treatment and the treatment of individual $i$ does not impact the potential outcome of individual $i' \neq i$. Assumption (A2) states that the treatment is completely randomized and assumption (A3) states that there is a non-zero probability that individual $i$ is assigned to either treatment or control. Assumptions (A1)-(A3) are usually satisfied by a RCT and consequently, $\tau$ can be identified by the well-known, difference-in-means formula, i.e., $\tau = E[Y_i \mid A_i = 1] - E[Y_{i} \mid A_i = 0]$. In particular, we review two consistent estimators of the ATE in RCTs, the difference-in-means estimator $\widehat{\tau}_{\rm Diff}$ and the covariate-adjusted, doubly robust estimator $\widehat{\tau}_{\rm Cov}$.
	\begin{align}
		\widehat{\tau}_{\rm Diff}  &= \left(\frac{\sum_{i=1}^{n} Y_{i} A_i}{\sum_{i=1}^{n} A_i} \right)  - \left(\frac{\sum_{i=1}^{n} Y_{i} (1-A_i)}{\sum_{i=1}^{n} 1-A_i} \right) \\
		\widehat{\tau}_{\rm Cov} &=  \left(\frac{ \sum_{i=1}^{n} (Y_{i} - f_1(X_i)) A_i}{ \sum_{i=1} A_i} + \frac{1}{n} \sum_{i=1}^{n} f_{1}(X_i) \right) - \left( \frac{ \sum_{i=1}^{n} (Y_{i} - f_0(X_i))(1- A_i)}{ \sum_{i=1} 1-A_i} + \frac{1}{n} \sum_{i=1}^{n} f_{0}(X_i)\right)
	\end{align}
	Here, $f_1(x) = E[Y_i \mid A_i = 1, X_i = x]$ and $f_0(x) = E[Y_i \mid A_i = 0, X_i = x]$ are outcome regression models for treatment and control groups, respectively. The difference-in-means estimator does not require an outcome model and does not adjust for covariates $X_i$. The doubly robust estimator is an augmented version of the difference-in-means estimator that adjusts for covariates and, in a randomized experiment, is consistent even if $f_1$ or $f_0$ are mis-specified. But, if $f_1$ and $f_0$ are correctly specified, $\widehat{\tau}_{\rm Cov}$ is more efficient than $\widehat{\tau}_{\rm Diff}$. %But, empirically even if $f_1$ and $f_0$ are mis-specified, $\widehat{\tau}_{dr}$ tends to have smaller standard errors than $\widehat{\tau}_{w}$. 
	Finally, both estimators are asymptotically Normal with mean $\tau$ and their standard errors can usually be estimated by a sandwich formula or the bootstrap \citep{efron}. For additional discussions, see \citet{Lunceford}, \citet{Zhang2008}, \cite{Tsiatis2008}, and \citet{BangRobins}.
	
	While these estimators have desirable statistical properties, from a data privacy perspective, the estimators require individuals' responses $Y_1,\ldots,Y_n$. If individuals are unwilling or apprehensive about sharing their responses to treatment or if they share a dishonest response due to reservations about their data privacy, the estimators may no longer be consistent. The next few sections talk about a formal way to ``privatize" $Y_i$ and how to use this privatized response to identify and estimate treatment effects on $Y_i$. Also, while we can also privatize the pre-treatment covariates $X_i$ in a similar fashion as the response, we leave them to take on any arbitrary values since we can identify the causal effect without $X_i$. Instead, we'll primarily use $X_i$ to gain efficiency; see Sections \ref{sec:ident} and \ref{DRest} for details.
	
	\subsection{Privatizing Responses with Differential Privacy and Forced Randomized Response (FRR)}\label{dpdef}
	As before, let $Y_i$ be the original response that individual $i$ wishes to keep private and
	let $\widetilde{Y}_i$ be the ``privatized'' version of the original response that the investigator actually collects from an experiment. The private response $\widetilde{Y}_i$ is generated by a ``privatizing'' function $\mathcal{M}$ where $\mathcal{M}$ takes the original response $Y_i$ as the input and returns $\widetilde{Y}_i$ as the output. Some trivial examples of $\mathcal{M}$ include: (1) a constant function where for any $Y_i \in \mathcal{D}$, $\mathcal{M}(Y_i) = 0$ and (2) an identity function where for any $Y_i \in \mathcal{D}$, $\mathcal{M}(Y_i) = Y_i$. Intuitively, the first example is more privacy-preserving than the second example in that the constant function always produces the same value $0$ for every individual, making it impossible for an investigator to recover the original, sensitive response $Y_i$ from $\widetilde{Y}_i$. But, the first example's $\mathcal{M}$ makes the ATE unidentifiable since everyone in the study has the same outcome, irrespective of the treatment.
	In contrast, in the second example with the identify function,  identification of the ATE is possible, but it is not privacy-preserving since the investigator sees the original, sensitive response $Y_i$ of individual $i$, i.e., $\widetilde{Y}_i = Y_i$. More broadly, \citet{Dwork&Roth} argued that most non-randomized functions $\mathcal{M}$ are inadequate to simultaneously preserve privacy and allow estimation and consequently, \citet{Dwork2006} proposed differential privacy, a family of non-deterministic $\mathcal{M}$s that take an input and stochastically generates an output. For example, if the input is individual $i$'s original response $Y_i$, $\mathcal{M}$ may return $Y_i$ plus some stochastic noise generated by $\mathcal{M}$, say $Y_i$ plus a random value from a Laplace distribution. The investigator specifies $\mathcal{M}$, carefully choosing how ``random'' $\mathcal{M}$ should be in order to balance the need for data privacy and the need to estimate scientifically meaningful quantities; too much randomness would mean that the ATE becomes ``less identifiable'' while too little randomness would mean that the individual's data is less private. The amount of this randomness is measured by the privacy loss parameter $\epsilon$, $\epsilon \in [0,\infty)$, and we say a random map $\mathcal{M}$ is $(\epsilon,0)$-differentially private if changing the input to $\mathcal{M}$ only changes its output by a factor based on $\epsilon$; see Definition \ref{def:M}.
	\begin{definition}[Differential privacy \citep{Dwork&Roth}] \label{def:M}
		A randomized map $\mathcal{M}$ is $(\epsilon, 0)$-differentially private for $\epsilon \geq 0$, if for all pairs of $Y, Y' \in \mathcal{D}$ with $Y \neq Y'$ and for any measurable set $S$ in the range of $\mathcal{M}$, we have $\Pr[\mathcal{M}(Y)\in S] \leq \exp(\epsilon) \cdot \Pr[\mathcal{M}(Y')\in S]$. 
	\end{definition}
\noindent Lower $\epsilon$  values imply that $\mathcal{M}$ becomes more privacy preserving; in the extreme case where $\epsilon =0$ and there is no loss in privacy, the output from $\mathcal{M}$ is statistically indistinguishable for any pair of inputs $Y$ and $Y'$ and estimation of treatment effects would be impossible. An example of such $\mathcal{M}$ would be a fair coin toss where, regardless of the original input $Y_i$, $\Pr(\mathcal{M}(Y_i) = 1) = \Pr(\mathcal{M}(Y_i) = 0) = 1/2$, i.e., individual $i$'s privatized response $\widetilde{Y}_i$ would be a result of a coin toss. On the other hand, by setting $\epsilon > 0$ , $\mathcal{M}$ allows some ``signal'' from the private outcome $Y_i$ to be passed onto the privatized outcome $\widetilde{Y}_{i}$ so that treatment effects can be estimated. Some values of $(\epsilon,0) \text{-differential privacy}$ that are used in practice are $(4,0)$ \citep{Apple}, $(3,0)$ \citep{Qinetal2016}, or approximately $(1.5,0)$ \citep{fanti2015building}. We also remark that unlike the usual definition of differential privacy which considers privacy of databases containing records from multiple individuals, Definition \ref{def:M} is specific to a single entry in a database and as such, is a local definition of differential privacy \citep{Dwork&Roth,Duchi}.	
	
	The function $\mathcal{M}$ that we will use in our proposed design is based on the forced randomized response (FRR) of \citet{Fox}. %ee \textcolor{red}{Section FILL} of supplementary materials for additional examples of $\mathcal{M}$. 
	Broadly speaking, in an FRR, individuals use a randomization device, typically a six-sided die, and based on the results of the randomization, some are instructed or ``forced'' to give a specific type of response, say $0$ or $1$, regardless of their original response $Y_i$, while others are instructed to give their original response $Y_i$. %Where the question of interest is usually asked in such a manner that a ``yes" response means possessing a sensitive trait or acknowledging participation in a sensitive behaviour. 
	For example, if individual $i's$ die roll is on 1, individual $i$ is instructed to report a $0$ to the investigator, i.e., $\widetilde{Y}_{i} = 0$, and if the die roll is on 6, individual $i$ is instructed to report a $1$ to the investigator, i.e., $\widetilde{Y}_{i} = 1$. If the die roll is on anything  but 1 or 6, individual $i$ is instructed to provide the original, true, response $\widetilde{Y}_{i} = Y_i$. A bit more formally, if $P_i \in \{0,1,2\}$ represent different instructions where $P_i = 0$ represents `report 0', $P_i = 1$ represents `report 1', and $P_i = 2$ represents `report original response', with probabilities $r_0, r_1$ and $1-r_0-r_1$, respectively, an FRR map $\mathcal{M}$ is defined as,
	{\small	\[
	\widetilde{Y}_i = \mathcal{M}(Y_i) = \begin{cases}
		0 & \text{if $P_i = 0$; $\Pr(P_i = 0) = r_0$} \\
		1 & \text{if $P_i = 1$;  $\Pr(P_i = 1) = r_1$}\\
		Y_i & \text{if $P_i = 2$; $\Pr(P_i=2) = 1-r_0 - r_1.$}
	\end{cases}
	\]}
	A critical part of a FRR is that investigators are unaware of the result of the die roll; only the participant knows the result of the die toss. Hence, investigators have no idea if $\widetilde{Y}_i$ from individual $i$ is the true $Y_i$ or one of the forced responses $0$ or $1$. To put it differently, a FRR protects the privacy of individual $i$'s response by plausible deniability of any particular response. Nevertheless, investigators can choose the die probability, represented by $r_0$ and $r_1$, and these values affect both the privacy loss $\epsilon$ and efficiency of $\tau$; in Section \ref{sec:DP}, we present a formula relating FRR parameters $r_0,r_1$ with the privacy loss parameter $\epsilon$ in differential privacy. For simplicity, we will collapse $r_{1} = r_{0} = r$ going forward so that there is one parameter $r$ parametrizing the privacy-preserving $\mathcal{M}$.

	\subsection{Noncompliance to Experimental Protocol and Cheaters }\label{sec:cheat_eg}

	Suppose an individual is generally wary of sharing their response to treatment and does not trust the privacy-preserving nature of $\mathcal{M}$. For example, a participant may feel like the die in a FRR is rigged and produce a response $\widetilde{Y}_i$ that deviates from the original FRR protocol. Or, a participant, after being instructed to report the true response $Y_i$ from the die roll, may feel uncomfortable doing so and instead report the opposite, say $\widetilde{Y}_{i} = 1 - Y_i$, to the investigator. These are instances of noncompliance to the experimental protocol and our goal is to still estimate causal effects in the presence of it.
	
	To achieve this goal, we use the concept of ``cheaters" by \citet{Clark&Desharnais} in psychometric testing. Broadly speaking, cheaters are those who deviate from the experimental protocol laid out by investigators whereas non-cheaters/honest individuals are those who follow the experimental protocol. For example, an individual who reports `0' (i.e., $\widetilde{Y} = 0$) even though the FRR prompt was to report `1' is a cheater; see Table 1 in supplementary materials for more examples. %In particular, participants who follow the FRR map $\mathcal{M}$ are non-cheaters and those who deviate from the FRR by providing their own responses are classified as cheaters. If every participant cheats, there is no hope of identifying a causal effect
	\citet{Clark&Desharnais} showed that if all cheaters are assumed to generate the same response to the investigators, say all cheaters only report $\widetilde{Y}_i = 0$, we can detect the presence of cheaters by (a) randomly splitting the study sample into two pieces, say $n$ individuals are split into samples of size $n_1$ and $n_2$, $n_1 +n_2 = n$, and (b) comparing appropriate statistics between the two subsamples. Our work extends this idea of using sample splitting to detect cheaters by relaxing the requirement that all cheaters generate the same response. We achieve this by using a different statistic to compare between the two subsamples, namely a variant of the compliance rate in instrumental variables \citep{AngristImbensRubin}; see Section \ref{sec:ident} for details.

	\section{Robust, Private Randomized Controlled  Trial}\label{design&inference}
	\subsection{Protocol} \label{exp_protocol}
\begin{figure}
		\begin{center}
			\includegraphics[width=\textwidth,height=\textheight,keepaspectratio]{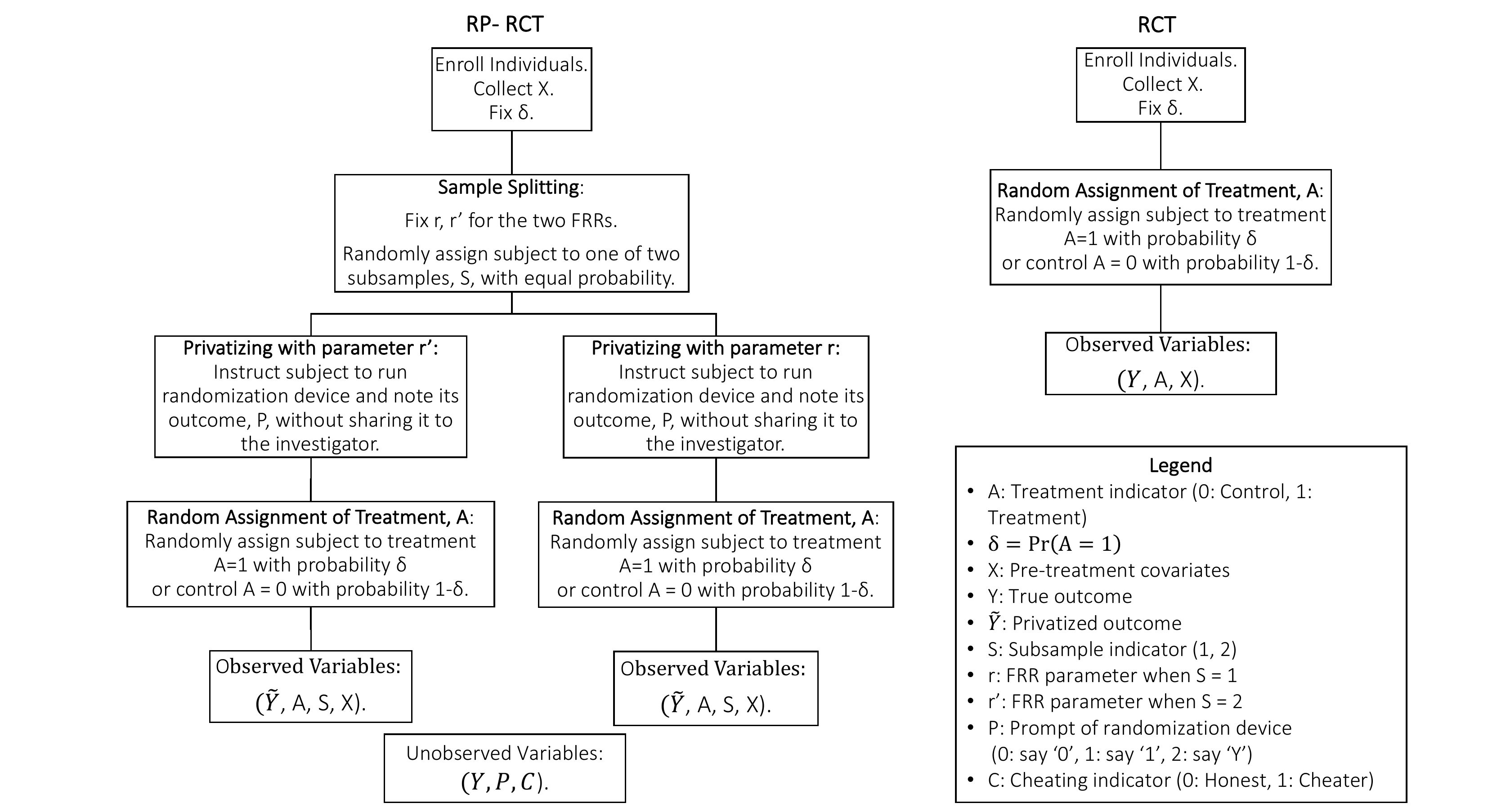}
		\end{center}
		\caption{Protocol for a Robust, Differentially Private Randomized Controlled Trial (RP-RCT) (left) Versus a Traditional RCT (right). Compared to a RCT, a RC-PCT adds two additional steps, sample splitting and FRR. Sample splitting is used to make our design robust to responses from cheaters and FRR is used to privatize responses.}  \label{fig:P-RCT}
	\end{figure}

	Combining the aforementioned ideas on differential privacy and cheaters, we propose a robust and differentially private RCT,  which we call a RP-RCT and is laid out in Figure \ref{fig:P-RCT}. In a RP-RCT, the investigator specifies (i) the treatment assignment probability $\delta = \Pr(A_i = 1)$ that is away from $0$ and $1$ and (ii) two slightly different FRR maps $\mathcal{M}, \mathcal{M'}$ parametrized by $r, r' \in [0,0.5)$, $r \neq r'$. The output of a RP-RCT is the individual $i$'s (a) treatment assignment $A_i$, (b) privatized response $\widetilde{Y}_{i}$ that may be contaminated due to cheaters, and (c) $S_{i} \in \{1,2\}$ indicating which of the two sub-samples individual $i$ belonged to or, equivalently, which FRR they were assigned to. The investigator does not observe the original response $Y_i$, the cheating status of individual i, denoted as $C_i$ where $C_i =1 $ if $i$ is a cheater and $0$ otherwise, and the result of the randomization device behind the FRR, $P_i$; these are denoted as unobserved variables in Figure \ref{fig:P-RCT}. In particular, the variable $C_i$ can be thought of as a latent characteristic of individual $i$ that is never observed by the investigator and the investigator only sees $\widetilde{Y}_{i}$ without knowing whether it came from a cheater (i.e $C_i = 1$) or not (i.e., $C_i = 0$).

	We make some remarks about the experimental protocol. First, compared to a traditional RCT, a RP-RCT adds two additional steps, the sample splitting step to be robust against cheaters and the FRR step to privatize the study participant's response. Here, for simplicity, the sample splitting step creates two equal, random, sub-samples of individuals, but this can be relaxed at the expense of additional notation. Second, a RP-RCT satisfies assumptions (A2) and (A3) because the treatment $A_i$ is still randomly assigned to individuals with probability $\delta$. Also, so long as the treatment is well-defined and does not cause interference (i.e., does not violate SUTVA), all assumptions (A1)-(A3) are satisfied by a RP-RCT. Third, while we only consider the FRR as our $\mathcal{M}$, we can replace $\mathcal{M}$ with another differentially private algorithm. Fourth, because the treatment $A_i$ is randomized, the pre-treatment covariates $X_i$ can take on any value (e.g., missing, censored, etc.) and it won't impact the identification strategy.
	
	\subsection{Differential Privacy Guarantee}\label{sec:DP}
	The following theorem shows that among non-cheaters, the privatized response $\widetilde{Y}_i$ generated from a RP-RCT is $(\ln\left(\frac{2}{r+r'}-1 \right), 0)$-differentially private.
	\begin{theorem}[Differential Privacy of RP-RCT] \label{thm:dp}
		Consider a non-cheater's true response $Y_i$. Then, a RP-RCT which generates his/her privatized response $\widetilde{Y}_{i}$ is $(\ln\left(\frac{2}{r+r'}-1 \right), 0)$ differentially private. 
	\end{theorem}
	\noindent In words, Theorem \ref{thm:dp} states that regardless of the non-cheater's true response $Y_i$, the privatized outcome generated from a RP-RCT, $\widetilde{Y}_{i}$ would be private up to some privacy loss parameter. The exact privacy loss depends on the FRR parameters $r, r'$ and investigators can choose $r,r'$ to achieve the desired level of privacy loss. 
	Also, Theorem \ref{thm:dp} does not make any claims about differential privacy for cheaters. This is because cheaters can choose to report any response $\widetilde{Y}_{i}$ that may or may not be differentially private. For example, if a cheater provides the result of a random coin flip as $\widetilde{Y}_{i}$ irrespective of his/her true response $Y_i$, then his/her response achieves perfect differential privacy, i.e., $\epsilon=0$. However, if a cheater always provides the opposite of his/her true response to potentially hide his/her true response, i.e., $\widetilde{Y}_{i} = 1-Y_i$, then his/her data, despite their best intentions, is never differentially private. Without making an assumption about cheaters' behaviors, we cannot make any guarantees about their responses' differential privacy. %Relatedly, the differential privacy guarantee among non-cheaters is not affected by the behavior of cheaters or the proportion of cheaters.

	\subsection{Identification of the ATE} \label{sec:ident}
	To lay out the identification strategy with a RP-RCT, we assume that the RP-RCT satisfies the following assumptions.
	\begin{itemize}
		\item[(A4)] Random sample splitting: $S_i \perp A_i, Y_i, Y_i(1), Y_{i}(0), X_i, C_i$ with $\Pr(S_i = 1) = \Pr(S_i=2) = 0.5$.
		\item[(A5)] Randomization in FRR: $P_i \perp A_i, Y_i, Y_i(1), Y_{i}(0), X_i, C_i$ with $r, r' \in [0,0.5)$ and $r \neq r'$
		\item[(A6)] Extended randomization of treatment: $A_i \perp S_i, P_i, Y_i(1), Y_{i}(0), X_i, C_i$.
	\end{itemize}
	Assumption (A4) states that the two subsamples in a RP-RCT were split randomly. Assumption (A5) states that the randomization device in the FRR (i.e., the die roll) is random and the two FRRs used in the two subsamples are different. Assumption (A6) is a re-iteration of the treatment randomization assumption (A2), except we now include the new variables introduced as part of a RP-RCT: $S_i, P_i$ and $C_i$. Note that Assumptions (A4)-(A6) are satisfied by the design of a RP-RCT.

	Let $\lambda \in [0,1)$ be the proportion of cheaters in the population, i.e., $\Pr(C_i = 1) = \lambda$. We now state assumptions about the cheater status $C_i$. While likely in some settings, these assumptions may not always be satisfied by the design of a RP-RCT.
	
	\begin{itemize}
		\item[(A7)] Cheater's response: $ \widetilde{Y}_i \perp A_i \mid S_i, P_i, C_i = 1$. %previously made assumption, added back to list
		%\item[(A7)] Cheating status: $C_i\perp Y_i(1), Y_i(0)$ 
		\item[(A8)] Proportion of non-cheaters: $0 < \Pr(C_i = 0)= 1-\lambda$.
	\end{itemize}
	Assumption (A7) states that a cheater gives the same response to the investigator regardless of whether he/she was randomized to treatment or control; note that assumption (A7) does not say that all cheaters produce the same response (i.e., the assumption underlying \cite{Clark&Desharnais}). Assumption (A7) is plausible if the treatment assignment $A_i$ is blinded so that the participant does not know which treatment he/she is receiving and thus, cannot use this information to change his/her final response $\widetilde{Y}_{i}$ to the investigator. Also, assumption (A7) still allows a cheater to use his/her original response $Y_i$, potential outcomes $Y_i(1), Y_i(0)$, or pre-treatment characteristics $X_i$ to tailor his/her private response $\widetilde{Y}_i$. For example, if a cheater reports a constant value, say $\widetilde{Y}_{i}=0$, or the opposite of his true response, say $\widetilde{Y}_{i} = 1 - Y_i$, assumption (A7) will still hold. Or if some cheaters' private response depends on unmeasured, pre-treatment characteristics, say, cheaters who are more privacy-conscious report $\widetilde{Y}_{i} = 0$ while less privacy-conscious cheaters may report a mixture of $0$ or $Y_i$, assumption (A7) will hold. However, if a cheater uses the treatment assignment to change his/her response $\widetilde{Y}_i$, say the cheater would report $\widetilde{Y}_i = 1$ if they receive treatment and $\widetilde{Y}_{i} = 0$ if they receive control, assumption (A7) is violated. Assumption (A8) states that not all participants in the study are cheaters. If assumption (A8) does not hold and every participant is a cheater, we cannot identify any treatment effect. Also, in Section \ref{sec:inference}, we present a way to assess assumption (A8) with the observed data by estimating the proportion of cheaters. Overall, so long as the treatment is blinded and there is at least one non-cheater, a RP-RCT plausibly satisfies assumptions (A1)-(A8) by design.

	We now show that the data from a RP-RCT can identify the ATE among non-cheaters.
	\begin{theorem}[Identification of the ATE Under RP-RCT] \label{thm:ident} Consider a RP-RCT with FRR parameters $r,r' \in [0,0.5)$ set by the investigator and the observed data is $(\widetilde{Y}_{i}, A_{i}, S_{i}, X_i)$. Under assumptions (A1)-(A8), the ATE among non-cheaters, denoted as $\tau_{\rm H} = E[Y_i(1) - Y_{i}(0) \mid C_i = 0] $, is identified from data via
		
		\begin{equation}
			\tau_{\rm H} = \frac{E[\widetilde{Y}_i|A_i=1]-E[\widetilde{Y}_i|A_i=0]}{(1-\lambda)(1-r-r^{\prime})}.
		\end{equation}
		
		Additionally, the proportion of cheaters $\lambda$ is identified via
		\begin{equation} \label{eq:lambda_ident}
			\lambda = 1 - \frac{1-2r'}{r-r'}E(\widetilde{Y}_i \mid S_i = 1) + \frac{1-2r}{r-r'}E(\widetilde{Y}_i \mid S_i = 2).
		\end{equation}
	\end{theorem}
\noindent Theorem \ref{thm:ident} shows that our new design can identify the ATE among non-cheaters by taking the difference in the averages of the privatized outcomes $\widetilde{Y}_{i}$ among treated and control units, re-weighed by the FRR parameters $r, r'$ and the proportion of cheaters $\lambda$. If there are no cheaters in the population, Theorem \ref{thm:ident} implies $\tau = \tau_{\rm H}$ and we can identify the ATE for the entire population. In contrast, if everyone is a cheater, we cannot identify the treatment effect; intuitively, if everyone is a cheater, disregards the FRR, and reports $\widetilde{Y}_{i} = 0$, it would be impossible to know the effect of the treatment on the response. Generalizing this intuition, we can only identify the subpopulation of individuals who are non-cheaters, even if the investigator does not know who are cheaters or non-cheaters. We remark that this result is similar in spirit to the local average treatment effect in the noncompliance literature \citep{AngristImbensRubin} where under noncompliance, only the subpopulation of compliers can be identified from data.
	
	Theorem \ref{thm:ident} also shows that we can identify the proportion of cheaters. While the exact formula is complex, roughly, we measure the excess proportion of privatized outcomes had everyone followed the FRR  and use additional moment conditions generated from sample splitting to identify $\lambda$; see Section B of the supplementary materials for details. %i.e, the proportion of 1s that are from cheaters not following the FRR prompt. 
	
	Overall, Theorems \ref{thm:dp} and  \ref{thm:ident} show that we can identify the treatment effect with privatized outcomes $\widetilde{Y}_{i}$, some of which may be contaminated by cheaters. The proposed design has some key parameters, $r$ and $r'$, that govern the privacy loss parameter $\epsilon$. They also affect estimation and testing of $\tau$, which we discuss below.
	
	\subsection{Difference-in-Means Estimator of the ATE} \label{sec:inference}
	Using the identification result in Section \ref{sec:ident}, we can construct estimators of $\tau_H$ by replacing the expectations in Theorem \ref{thm:ident} with their sample counterparts, i.e., %This section explores various approaches Section \ref{sec:consMLE} also discusses estimation of $\pi, \gamma, \lambda$ based on constrained MLE in order to improve its finite-sample performance.
{\small 
	\begin{align}
		\widehat{\tau}_{\rm H, Diff} &= \frac{\widehat E[\widetilde{Y}_i|A_i=1]-\widehat E[\widetilde{Y}_1|A_i=0]}{(1-\widehat \lambda)(1-r-r^{\prime})} =\frac{1}{(1- \widehat\lambda)(1-r-r^{\prime})}  \left(\frac{\sum_{i=1}^{n} \widetilde{Y}_{i} A_{i}}{\sum_{i=1}^{n} A_{i}}-\frac{\sum_{i=1}^{n} \widetilde{Y}_{i} (1-A_{i})}{\sum_{i=1}^{n} (1-A_{i})}\right).
	\end{align}}
	\noindent Note that the estimator $\widehat{\tau}_{\rm H, Diff}$  can be thought of as an extension of the simple difference-in-means estimator in a RCT re-weighted by the estimated proportion of non-cheaters and the privacy-preserving map, i.e., $(1-\widehat\lambda)(1-r-r')$. Also, an estimate of the proportion of cheaters $\widehat{\lambda}$ can be obtained by replacing the expectations in Theorem \ref{thm:ident} with their sample counterparts, i.e.,
	\begin{equation} \label{eq:lambda_hat}
		\widehat{\lambda} = 1 - \frac{1-2r'}{r-r'}\widehat{E}(\widetilde{Y}_i \mid S_i = 1) + \frac{1-2r}{r-r'}\widehat{E}(\widetilde{Y}_i \mid S_i = 2).
	\end{equation} 
Also, $\widehat{\lambda}$ is the maximum likelihood estimate of $\lambda$; see Section B.3.1 of the supplementary materials for details. Also, in small samples, $\widehat{\lambda}$ may exceed the bound of $0$ and $1$, especially if $r$ and $r'$ are close to each other. In this case, we follow \citet{Clark&Desharnais}'s advice where we evaluate the likelihood of the estimated $\widehat{\lambda}$ and the boundary points $0,1$ and pick the value of $\lambda$ corresponding to the highest likelihood.

	%Let $y_1, m_1$ and $y_2, m_2$ be the number of `yes' responses and sample sizes under subsample-1 and subsample-2, respectively. 
\noindent Theorem \ref{thm:IPW} shows that $	\widehat{\tau}_{\rm H, Diff}$ is a consistent and asymptotically Normal estimator of $\tau_{\rm H}$.
	
	\begin{theorem}[Asymptotic Properties of $\widehat{\tau}_{\rm H, Diff}$] \label{thm:IPW}
		Suppose the observed data $(\widetilde{Y}_i, A_i, S_i)$ is i.i.d. and generated from a RP-RCT. Then, we have $\sqrt{n}(\widehat\tau_{\rm H,Diff} -\tau_{\rm H}) \xrightarrow{d} N(0,V_{\rm H, Diff})$ with
		{\small 	
			$$V_{\rm H, Diff}=  \frac{1}{(1-r-r')^2(1-\lambda)^2}  \left[ \frac{Var(\widetilde Y_i \mid A_i = 1)}{\delta} + \frac{Var(\widetilde Y_i \mid A_i = 0)}{1-\delta}\right]+ \tau_H^2 \left[2+ \frac{Var(\widehat{\lambda})}{(1-\lambda)^2}\right].$$}
		
		Also, $V_{\rm H, Diff}$ can be consistently estimated by,
		{\small 	
			$$\widehat{V}_{\rm H, Diff} = \frac{1}{(1-r-r')^2(1-\widehat\lambda)^2}\left[ \frac{\widehat{Var}(\widetilde Y_i \mid A_i = 1)}{\frac{1}{n}\sum_{i=1}^{n} A_{i}} + \frac{\widehat{Var}(\widetilde Y_i \mid A_i = 0)}{1-\frac{1}{n}\sum_{i=1}^{n}A_{i}}\right] +\widehat{\tau}_{H,Diff}^2 \left[2+ \frac{\widehat{Var}(\lambda_i)}{(1-\widehat\lambda)^2}\right],$$} where, $\widehat{Var}(\widetilde Y_i \mid A_i = j) = \widehat E(\widetilde Y_i \mid A_i = j)\left[1- \widehat E(\widetilde Y_i \mid A_i = j)\right]$ for $j \in \{0,1\}$, 
		{\small 
			$$\widehat{\rm Var}(\widehat{\lambda}) = \frac{1}{(r-r')^2}  \left[ (1-2r')^2\frac{\widehat{Var}(\widetilde{Y_i}\mid S_i = 1) }{\frac{1}{n}\sum_{i=1}^{n} \mathbf{1}(S_i = 1)} + (1-2r)^2 \frac{\widehat{Var}(\widetilde{Y_i}\mid S_i = 2)}{\frac{1}{n}\sum_{i=1}^{n} \mathbf{1}(S_i = 2)} \right],$$}
		
		and, $\widehat{Var}(\widetilde Y_i \mid S_i = k) = \widehat E(\widetilde Y_i \mid S_i = j)\left[1- \widehat E(\widetilde Y_i \mid S_i = j)\right]$ for $k \in \{0,1\}$.
		
	\end{theorem}
	\noindent Theorem \ref{thm:IPW} can be used as a basis to construct $1-\alpha$ confidence intervals and p-values for testing the null hypothesis $H_0: \tau_H = \tau_0$. For example, a Wald-style test for $H_0$ would be $t = \sqrt{n}(\widehat \tau_{\rm H, Diff}-\tau_0)\widehat{V}_{\rm H, Diff}^{-1/2}$ and one would reject the null in favor of a two-sided alternative at level $\alpha$ if $|t|$ exceed $z_{1-\alpha/2}$, where $z_{1-\alpha/2}$ is the $1-\alpha/2$ quantile of the standard Normal distribution. We can also construct a Wald-based $(1- \alpha) \times 100\%$  level two-sided confidence interval for $\tau_H$ as $(\widehat{\tau}_{\rm H, Diff}-z_{1-\alpha / 2}\sqrt{\widehat{V}_{\rm H, Diff}/n},  \widehat{\tau}_{\rm H, Diff}+z_{1-\alpha / 2} \sqrt{\widehat{V}_{\rm H, Diff}/n})$. Note that similar to the usual difference in means estimator in Section \ref{sec:notation}, practitioners could also use the bootstrap to estimate the standard error of $\widehat{\tau}_{\rm H, Diff}$ and its associated confidence interval.

	\subsection{Cost of Data Privacy in RP-RCT} \label{sec:Power}
	In this section, we compare a RP-RCT to a traditional RCT as a way to study the cost of guaranteeing differential privacy on statistical efficiency. To begin, suppose $\lambda = 0$ so that both $\widehat{\tau}_{\rm H, Diff}$ and $\widehat{\tau}_{\rm Diff}$ estimate the same parameter $\tau$. The following theorem shows that $\widehat{\tau}_{\rm H, Diff}$ is never as asymptotically efficient as $\widehat{\tau}_{\rm Diff}$; in short, there is a statistical cost of using a RP-RCT to guarantee differential privacy of an individual's response. 
	\begin{theorem} \label{thm:Efficiency} [Relative Asymptotic Efficiency of $\widehat{\tau}_{\rm H, Diff}$ and $\widehat{\tau}_{\rm Diff}$] \label{thm:rel_ipw} For any $\epsilon \in (0, \infty)$ and $\lambda = 0$,  we have,
	\begin{align*}
	\mathcal{E}(\widehat{\tau}_{\rm H, Diff}, \widehat{\tau}_{\rm Diff}) & =  \frac{(1-\delta)\tau_1(1-\tau_1) + \delta \tau_0(1-\tau_0)}{(1-\delta) \left[\frac{1}{\exp(\epsilon)-1} + \tau_1\right]\left[\frac{1}{1-\exp(-\epsilon)} - \tau_1\right] + \delta \left[\frac{1}{\exp(\epsilon)-1} + \tau_0\right]\left[\frac{1}{1-\exp(-\epsilon)} - \tau_0\right]},
	\end{align*}

		where, $\tau_0 = E(Y_i \mid A_i = 0),\tau_1 = E(Y_i \mid A_i = 1)$, and $\mathcal{E}(\widehat{\tau}_{\rm H, Diff}, \widehat{\tau}_{\rm Diff}) = \frac{{\rm Var}(\widehat{\tau}_{\rm Diff})}{{\rm Var}(\widehat{\tau}_{\rm H, Diff})}$. Also, $\mathcal{E}(\widehat{\tau}_{\rm H, Diff}, \widehat{\tau}_{\rm Diff})$ is monotonically increasing in $ \epsilon$.
	\end{theorem}

	The relative efficiency between the two estimators, measured by ${\rm Var}(\widehat{\tau}_{\rm Diff})/{\rm Var}(\widehat{\tau}_{\rm H,Diff})$, is determined by $\epsilon$, which in turn is governed by FRR parameters $r, r'$. In particular, as individuals' responses become less private by an increase in privacy loss $\epsilon$, the relative efficiency approaches $1$. In fact, only when the privacy loss approaches infinity, i.e., $\epsilon \to \infty$, do we have relative efficiency equaling $1$ and thus, $ \widehat{\tau}_{\rm H, Diff}$ will never be as efficient as  $ \widehat{\tau}_{\rm Diff}$ as long as we want to guarantee some amount of data privacy. Investigators can use the formula in Theorem 4 to assess what value of privacy loss, $\epsilon$, works best in their own studies. In particular, we recommend investigators specify $\epsilon$ based on (a) their tolerance for loss in efficiency at the expense of more privacy, (b) $\tau_0$ and $\tau_1$, which may be informed from subject-matter experts during the planning stage of the experiment, and/or (c) recommended data privacy standards from relevant regulatory bodies.

	\subsection{Covariate Adjustment and Doubly Robust Estimation}\label{DRest} 
	Similar to a RCT, suppose a RP-RCT collected pre-treatment covariates $X_i$, which may be missing, contaminated, and/or corrupted. We propose to use the pre-treatment covariates to develop a more efficient estimator without incurring additional bias by using a doubly robust estimator \citep{BangRobins}. Formally, let $\widetilde{f_0}(x_i)$ and $\widetilde{f_1}(x_i)$ be the postulated models for the true outcome regressions $E\left[\widetilde {Y_i} \mid A_i = 0, X_i = x_i\right]$ and $E\left[\widetilde {Y_i} \mid A_i = 1, X_i = x_i\right]$, respectively. For simplicity, we assume that these functions are fixed, but our result below holds if these functions were estimated at fast enough rates, say those based on parametric models. Consider the following estimator for $\tau_{\rm H}$, 
{\small	\begin{align}
		\widehat\tau_{\rm H, Cov} &=\frac{\frac{1}{n} \sum_{i=1}^{n}\left[\frac{\left(\widetilde{Y}_{i}-\widetilde{f}_{1}\left(x_{i}\right)\right) A_{i}}{\delta}+\widetilde{f}_{1}\left(x_{i}\right)\right]-\frac{1}{n} \sum_{i=1}^{n}\left[\frac{\left(\widetilde{Y}_{i}-\widetilde{f}_{0}\left(x_{i}\right)\right)\left(1-A_{i}\right)}{1-\delta}+\widetilde{f}_{0}\left(x_{i}\right)\right]}{(1-\widehat{\lambda})\left(1-r-r^{\prime}\right)}.
	\end{align}}
	The following theorem shows that $\widehat{\tau}_{\rm H, Cov}$ is a consistent, asymptotically Normal estimate of $\tau_H$ even if $\widetilde{f}_1$ and $\widetilde{f}_0$ are mis-specified.
	
	\begin{theorem}[Consistency and Asymptotic Normality of $\widehat \tau_{\rm H, Cov}$]\label{thm: DR} Suppose the same assumptions in Theorem \ref{thm:IPW} hold. Then, for any fixed working models of the privatized outcomes $\widetilde{f}_1(x)$ and $\widetilde{f}_0(x)$, we have $\sqrt{n}(\widehat{\tau}_{\rm H, Cov} -\tau_{\rm H}) \xrightarrow{d} N(0,V_{\rm H, Cov})$ where	
{\footnotesize		$$V_{\rm H, Cov}= \frac{1}{(1-r-r')^2 (1-\lambda)^2}\left[E\left[(\widetilde{f_1}-\widetilde{f_0})^2\right] + \frac{E\left[(\widetilde{Y}_{i}-\widetilde{f_1})^2\mid A_i = 1\right]}{\delta} + \frac{E\left[(\widetilde{Y}_{i}-\widetilde{f_0})^2\mid A_i = 0\right]}{(1-\delta)} \right]+ \tau_H^2 \left[1+ \frac{Var(\widehat{\lambda})}{(1-\lambda)^2}\right].$$}
Also, $V_{\rm H, Cov}$ can be consistently estimated by,

{\footnotesize	\begin{align*}
		\widehat V_{\rm H, Cov} &= \frac{1}{(1-r-r')^2 (1-\widehat\lambda)^2}\left[\widehat E\left[(\widetilde{f_1}-\widetilde{f_0})^2\right] + \frac{\widehat E\left[(\widetilde{Y}_{i}-\widetilde{f_1})^2\mid A_i = 1\right]}{\delta} + \frac{\widehat E\left[(\widetilde{Y}_{i}-\widetilde{f_0})^2\mid A_i = 0\right]}{(1-\delta)} \right]\\
		&+ \widehat\tau_{H,Cov}^2 \left[1+ \frac{\widehat Var(\widehat{\lambda})}{(1-\widehat \lambda)^2}\right].
\end{align*}}

	\end{theorem}
	\noindent Similar to Theorem \ref{thm:IPW}, Theorem \ref{thm: DR} can be used to construct confidence intervals and p-values for testing $H_0: \tau_H = \tau_0$. Also, in Section A.5 of the supplementary materials, we characterize the relative efficiency between the doubly robust estimators under a RP-RCT and a RCT. 
	\section{Evaluation of Online Statistics Courses at the University of Wisconsin-Madison During COVID-19} \label{sec:data}
	
	\subsection{Background and Motivation}
%A RP-RCT was used to evaluate different types of of online video lectures on statistics at the University of Wisconsin-Madison. Briefly,
In light of many courses shifting online during COVID-19, the Department of Statistics at the University of Wisconsin-Madison was interested in evaluating which type of online video lectures best aided conceptual understanding, information retention and problem-solving among students taking the Department's courses. Specifically, one of the questions of interest was whether instructor-present lecture videos (i.e., treatment) led to a better learning experience for students compared to instructor-absent online lectures (i.e., control); see Figure \ref{fig:instructions} for an example. Some prior works \citep{WILSON, Kizilcec} found no evidence that instructor-present lecture videos had a significant impact on student learning in terms of attention and comprehension. However, other works \citep{Pi&Hong, WANG2020103779} show the opposite, that instructor-present lecture videos enhanced student learning. Regardless, the participant data from these works are not publicly available due to the sensitive nature of students' educational data. Additionally, there was concern that students may be less willing to provide their true attention and retention rate in the courses when asked by the Department. For example, some students might not be comfortable admitting to not paying attention to online lectures and simply lie to the investigator, leading to potentially biased results.

	Compared to using a RCT, using a RP-RCT provides numerous remedies for the aforementioned issues. First, students' data is guaranteed to be differentially private, which may encourage more honest participation. Second, even if some students remain dishonest and are cheaters, a RP-RCT still provides a robust estimate of the treatment effect. Third, the data can be shared publicly and students' data privacy is still preserved via differential privacy; as mentioned earlier, the experimental protocol, including sharing students' response data, was approved by the Education and Social/Behavioral Science IRB of the University of Wisconsin - Madison.\\

	\subsection{Study Participants, Treatment Arms, and Outcomes}
	
	The study population consisted of students enrolled in introductory statistics classes at the University of Wisconsin - Madison during the Spring 2021 semester. %, specifically March of 2021 to May of 2021.
	%The experiment was conducted online % through an interactive, Qualtrics survey 
	%and 
	Electronic, informed consent was obtained from all participants before enrollment. Once a student gave consent, the study collected the following pre-treatment covariates: gender, race/ethnicity, year in college, major or field of study, prior subject matter knowledge, previous exposure to and grades in statistics/mathematics/computer science classes, self-rated interest in statistics, proficiency in English, amount of experience with video lectures in the past or current semester, and preference on online lecture format. Students had the option to not provide answers to any of the pre-treatment covariates. Afterwards, students are randomly placed into one of the two subgroups ($S_i=1$ or $S_i =2$) and within each subgroup, they are randomly assigned to treatment or control. The control arm had a narrated `Instructor Absent' (IA) video format where students see the lecture slides and hear an audio narration of the lecture from the instructor. The treatment arm was identical to the control arm except it used an  `Instructor Present' (IP) video format where the instructor's face was embedded in the upper-right corner of the lecture video. Both lecture videos were 13 minutes long and introduced identical statistical concepts, specifically about RCTs. We remark that RCTs were not covered in the classes in which the study was conducted. Also, the study used an FRR with $r_0 = 0, r_1 = 0.1040$ and another FRR with $r_0' = 0, r_1' = 0.1667$, resulting in a privacy loss of $\epsilon = 2$. This value was based of our own preference towards data privacy at the expense of efficiency where we were willing to tolerate roughly $50\%$ increase in standard error under a RP-RCT than a RCT, and consultation with the University's IRB, which, among other things, gave approval to release the student-level data to the public for future replication and analysis.
	
	 After students watched either the IA or IP lecture video, they were asked a series of questions, notably on four areas of student learning (i.e., Attention, Retention, Judgement of Learning and Comprehension) used by previous works \citep{Kizilcec,Pi&Hong, WANG2017, WILSON}. All four outcomes were `yes/no' questions; the exact wording of the questions is in Table \ref{tab:POs}. 

	For our working models for the privatized outcomes $\widetilde{f}_1$ and $\widetilde{f}_0$, we used logistic regression models that minimized the Akaike information criterion \citep{AIC}; see Section C of supplementary materials for additional details.

	\begin{table}[]
		\centering
		\begin{tabular}{ll}
			\hline
			\multicolumn{1}{c}{Primary Outcome} & \multicolumn{1}{c}{Questions} \\ \hline
			Attention & ``I found it hard to pay attention to the video." \\
			Retention & ``I was unable to recall the concepts while attempting the followup quiz." \\
			Judgement of Learning & ``I don't feel that I learnt a great deal by watching the video." \\
			Comprehension & ``I found the topic covered in the video to be hard." \\ \hline
		\end{tabular}
		\caption{Questions Concerning Four Areas of Student Learning. Students were asked to say ``yes'' or ``no'' if they agreed or disagreed with the statements above.}
		\label{tab:POs}
	\end{table}
	
\begin{figure}[h!]
		\centering
		\includegraphics[width=\textwidth,height=\textheight,keepaspectratio]{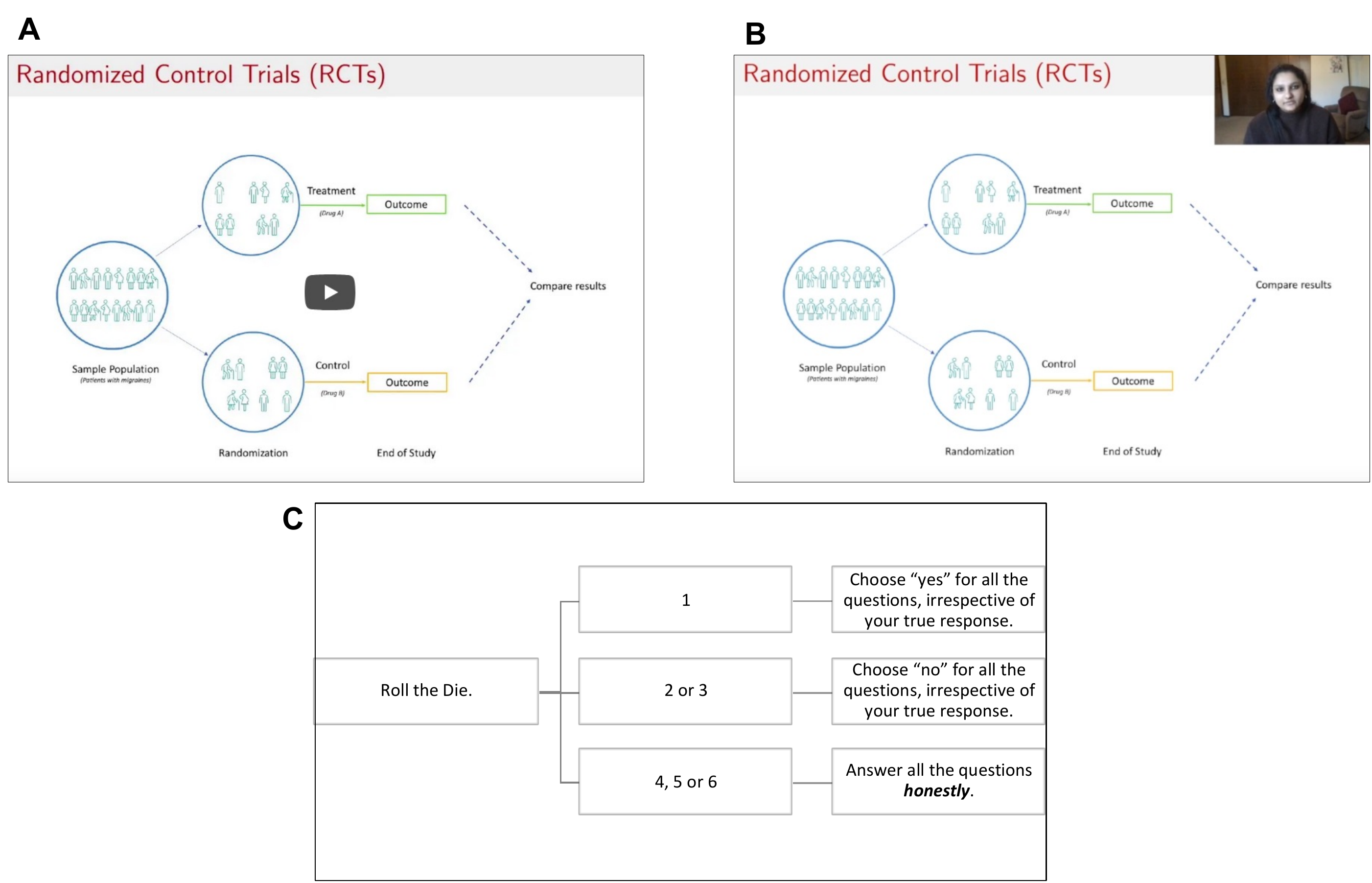}
		\caption{A Sample Slide of The instructor-Absent (Panel A) and Instructor-Present (Panel B) Video Lectures on RCTs. Panel C shows the instructions for the FRR.}
		\label{fig:instructions}
	\end{figure}

	We remark that as part of the RP-RCT protocol, students were prompted to `roll a die' using an online die roll (i.e., the FRR device). Students were allowed to roll the die only once and the resulting die roll was visible to the student only. Based on the outcome of the roll, students were asked to answer the four questions in Table \ref{tab:POs} using the FRR prompt presented in panel C of Figure \ref{fig:instructions}. Also, students must roll the die before being presented with the four questions.

	\subsection{Results}
	Table \ref{tab:classroom_results} presents the estimates of the average treatment effect amongst honest participants ($\tau_H$) for the four outcomes. Honest students indicated that it was easier to pay attention to instructor-present (IP) video lectures compared to instructor-absent (IA) video lectures ($\widehat{\tau}_{\rm H, Cov} = -0.316, 95\% \text{ CI: } [-0.621, -0.011]$). However, their ability to retain concepts was higher amongst students randomized to the instructor-absent (IA) video lectures ($\widehat{\tau}_{\rm H, Cov} = 0.395, 95\% \text{ CI: } [0.022,0.767]$). However, there was no significant difference between students who watched IP and IA video lectures in terms of Judgement of Learning and Comprehension. Also, Section C of the supplementary materials present covariate balance between the IP and IA subgroups and, as expected, we found no differences between the two groups in terms of their baseline pre-treatment covariates. Our results suggest that cheating varies with the question. For example, the estimated proportion of cheaters for the Attention and Judgement of Learning related questions were smaller ($24\%$ and $32\%$, respectively). However, they were larger for Retention and Comprehension questions ($40\%$ and $38\%$, respectively). These differences may suggest that students might be more apprehensive in sharing outcomes related to their learning abilities (i.e., Retention, Comprehension) than outcomes related to instruction (e.g., ability to engage students and transfer knowledge).

	\begin{table}[]
		\centering
		\begin{tabular}{lccc}
			\hline Primary Outcome & $\widehat{\lambda}$ & $\widehat{\tau}_{\mathrm{H}, \mathrm{Diff}}$  & $\widehat{\tau}_{\mathrm{H}, \mathrm{Cov}}$  \\
			\hline Attention & $0.240$ $(0.129)$ & $-0.163$ $(0.181)$  & $-0.316$ $(0.155)$ \\
		%& {$[0.026,0.453]$} & {$[-0.461,0.135]$} & {$[-0.575,-0.043]$} \\
			Retention & $0.400$ $(0.180)$ &  $\text{ } \text{ } 0.308$ $(0.226)$ & $\text{ } \text{ } 0.395$ $(0.190)$ \\
		%	& {$[0.103,0.696]$} & {$[-0.064,0.681]$} & {$[0.035,0.690]$} \\
			Judgement of Learning & $0.32	0$ $(0.148)$ & $\text{ } \text{ } 0.201$ $(0.213)$ & $\text{ } \text{ } 0.214$ $(0.198)$ \\
		%	& {$[0.105,0.594]$} & {$[-0.148,0.552]$} & {$[-0.111,0.540]$} \\
			Comprehension & $0.380$ $(0.199)$ & $-0.031$ $(0.314)$ & $-0.124$ $(0.297)$ \\
		%	& {$[0.051,0.708]$} & {$[-0.550,0.483]$} & {$[-0.613,0.364]$} \\
			\hline
		\end{tabular}
		\caption{Estimates of Treatment Effects on the Four Outcomes. Standard errors (s.e.) of $\widehat{\lambda}$ are estimated using 5000 bootstrapped samples.}% 90$\%$ confidence intervals are reported in square brackets.}
		\label{tab:classroom_results}
	\end{table}

	Our results largely agree with previous works on online video lectures. %Disparities between our study and previous works may be due to differences in processes, participants, lecture topics and instructors. 
	For example, our results and those by \cite{Kizilcec, Pi&Hong} and \cite{WANG2020103779} agree that IP lectures receive considerably more attention than IA lectures. Also, our findings on retention, judgement of learning, and comprehension match those in \cite{WANG2017, WANG2020103779,WILSON}. However, we remark that there is work \citep{Kizilcec} that suggests the opposite of what we find on retention. Finally, unlike these works, all of the student-level data and code is publicly available for replication and future analysis, especially if investigators want to combine this data with future datasets to boost power in related evaluations of online video lectures.

	\section{Conclusion and Discussion}
	
	We propose a new experimental design to evaluate the effectiveness of a program, policy, or a treatment on an outcome that may be sensitive, with a particular focus on online education programs where students' response data are often sensitive. Our design, a RP-RCT, has differential privacy guarantees while also allowing estimation of treatment effects. A RP-RCT also accommodates cheaters who may not trust the privacy-preserving nature of our design and provide arbitrary responses that may further protect their privacy. We provide two consistent, asymptotically Normal estimators, one of which allows for covariate adjustment. We also assess the trade-off between differential privacy and statistical efficiency. We conclude by using the RP-RCT to evaluate different types of online video lectures at the Department of Statistics at the University of Wisconsin-Madison and find that our results largely agree with existing results on online video lectures, while preserving students' data privacy and allowing sharing of this data for future replication.

	\bibliographystyle{Chicago}
	
	\bibliography{Bibliography-RPRCT}
\end{document}